\documentclass[journal=jpccck,manuscript=article]{achemso}
\setkeys{acs}{articletitle=true}
\usepackage[latin1]{inputenc}
\usepackage{chngpage}
\usepackage{amsmath}
\usepackage{amsfonts}
\usepackage{amssymb}
\usepackage{graphicx}
\usepackage{color}
\SectionNumbersOff

\def\<{\langle}
\def\>{\rangle}

\newcommand{\textgx}[1]{\textcolor{black}{#1}}

\title{Origin of Charge Separation at Organic Photovoltaic Heterojunctions: a Mesoscale Quantum Mechanical View}

\author{Mos\`{e} Casalegno}
\affiliation{Dipartimento di Chimica, Materiali e Ing. Chimica "G. Natta", Politecnico di Milano, Via L. Mancinelli 7, 20131 Milano, Italy}

\author{Raffaele Pastore}
\affiliation{CNR-SPIN, Via Cintia, 80126 Napoli, Italy}

\author{Julien Id\'{e}}
\affiliation{Laboratory for Chemistry of Novel Materials, University of Mons, Belgium}

\author{Riccardo Po}
\affiliation{Research Center for Renewable Energies $\&$ Environmental R$\&$D, Eni SpA, Via G. Fauser 4, 28100 Novara, Italy}

\author{Guido Raos}
\email{guido.raos@polimi.it}
\affiliation{Dipartimento di Chimica, Materiali e Ing. Chimica "G. Natta", Politecnico di Milano, Via L. Mancinelli 7, 20131 Milano, Italy}

\date{\today}

\begin{document}

\begin{abstract}
The high efficiency of charge generation within organic photovoltaic blends apparently contrasts with the strong "classical" attraction between newly formed electron-hole pairs. Several factors have been identified as possible facilitators of charge dissociation, such as quantum mechanical coherence and delocalization, structural and energetic disorder, built-in electric fields, nanoscale intermixing of the donor and acceptor components of the blends. Our mesoscale quantum-chemical model allows an unbiased assessment of their relative importance, through excited-state calculations on systems containing thousands of donor and acceptor sites. The results on several model heterojunctions confirm that the classical model severely overestimates the binding energy of the electron-hole pairs, \textgx{produced by vertical excitation from the electronic ground state}. Using physically sensible parameters for the individual materials, we find that the quantum mechanical energy difference between the lowest interfacial charge transfer states and the fully separated electron and hole is of the order of the thermal energy.
\end{abstract}

\maketitle

\section{Introduction}

Solar energy is by far the most abundant renewable energy source, and harvesting it to produce electricity and "solar fuels" (e.g., molecular hydrogen) seems to be the most promising route in the transition to an energetically sustainable future.\cite{Solar} Silicon-based solar cells are already making a significant impact on worldwide energy production, but other photovoltaic technologies are being actively researched for the medium and long term. 
Organic photovoltaic (OPV) devices represent one of the alternatives, which could be attractive for some applications in view of the wide availability of raw materials, low production costs and printability on mechanically flexible substrates. Their key component is a thin semiconducting active layer consisting of a blend of an electron-donor (D) and an electron-acceptor (A), which may be conjugated polymers and/or small molecules.\cite{GREGG,BHJ1,BHJ2,BHJ3} Today, OPV devices with 10\% power conversion efficiency (PCE) have been produced by several groups,\cite{Liu14,Zhao16} and a record 12\% PCE has been achieved with a ternary blend.\cite{12PCE} This has been possible thanks to a careful selection of the materials---synthetic possibilities are almost~\cite{Po15} limitless---and optimization of the blend structure and morphology by controlling the deposition methods and post-deposition treatments.\cite{BHJ3,Salleo13}

As early as 2009, the group of Lee and Heeger achieved near-100\% internal quantum efficiency (IQE) in "bulk heterojunction" cells having 6\% PCE, based on a low-bandgap donor copolymer and a fullerene-based acceptor.\cite{100IQE} 
IQE measurements are somewhat difficult and therefore they are not usually performed in experimental studies, but it seems likely that IQE's exceeding 90\% should be achieved in all the current state-of-the-art devices. 
Such high values imply that virtually every absorbed photon---the IQE is actually a function of their wavelength---is successfully converted into a negatively charged electron (transported through the A material to the cell's cathode) and a positively charged hole (transported through the D material to the cell's anode). In turn, this implies a near-100\% success in each of the processes which follow the formation of an exciton by photon absorption within either phase.
According to the conventional wisdom, these are the diffusion of the exciton to the D-A interface, its dissociation into a "bound" electron-hole pair or charge-transfer (CT) exciton, their separation into free charge carriers and the migration/collection of the latter at the electrodes. 
A truly remarkable result, which apparently defies simple "classical" explanations: The attractive interaction between a positive and a negative point charge at 1-2 nm distance in a medium with relative permittivity $\varepsilon_r$=3-4 is 0.2-0.5 eV, which is much greater than $k_BT$=0.025 eV at room temperature. This and other observations have prompted the suggestion that to understand organic photovoltaics it is essential to invoke general quantum mechanical principles of delocalization, coherence and uncertainty,\cite{Heeger13,Mukamel13} and that it might be possible to enhance the performance of OPV devices by properly harnessing them.\cite{Scholes16}

Spectroscopic experiments have been carried out with a range of methods, allowing the characterization of the relevant species---excitons and polaronic charge carriers, as well as interfacial CT states---with increasing detail.\cite{Durrant10,Barano12,Cahen13,revBK,revFFN}
Ultrafast pump-probe experiments showed that high-energy, "hot" CT states (CT$_n$) tend to dissociate faster than the lower-energy ones (CT$_1$).\cite{Baku12,Grancini13,Rossky13,Friend16} A higher rate of charge separation was linked to a higher degree of quantum mechanical coherence and delocalization and was assumed to translate into a higher overall dissociation efficiency.
However, other experiments\cite{EFISH13,OA13} could be interpreted more conventionally in terms of a slow (on the ps scale, in comparison with the fs scale of the previous ultrafast studies), diffusive dissociation of "classical" charge carriers.
Salleo, Neher and coworkers~\cite{Salleo14,Neher14} have reported that "cold" CT$_1$ states produced by direct, weakly allowed absorption from the ground state dissociate just as efficiently (or as inefficiently, depending on the D:A combination) as the higher energy ones.
On the theoretical front, charge photogeneration has been modelled by accurate excited-state or time-dependent calculations on few-molecule systems,\cite{Few14,Baum14,OP14,Aki16} microelectrostatic~\cite{MES09,Poelk14} or quantum chemical calculations\cite{TVAN13,SavoieFriend14,DAvino} on larger D:A aggregates produced by molecular dynamics simulations,\cite{Pao11,MES14} or Kinetic Monte Carlo (KMC)\cite{KMC1,KMC2} and Master Equation (ME)\cite{ME} simulations at the scale of whole OPV devices. These methods have complementary strengths and weaknesses, but overall it has proved difficult to combine them to provide a general, fully satisfactory answer to the long-standing question: "Why is exciton dissociation so efficient at the interface between a conjugated polymer and an electron acceptor?".\cite{Arkhi03}

To sum up, several candidates have been identified as likely "facilitators" of CT dissociation: built-in electric fields at D:A interfaces, delocalization of the excitons and of the charge carriers, high charge mobility, energetic variability and structural disorder, domain size and degree of intermixing of the D and A "phases", non-linearity or inhomogeneity of the dielectric medium, excess energy of the photogenerated excitons.\cite{Durrant10,Barano12,Cahen13,revBK,revFFN} All these factors seem have some importance, but probably not equally so. Besides, some of them are likely to be incompatible with each other (e.g., disorder and delocalization/mobility of the charges).

Here we present fresh theoretical insights based on our effective two-orbital quantum chemical model,~\cite{refCGQC} which provides a "minimal" but theoretically sound description of OPV materials. It is similar in spirit to those of Troisi,\cite{TroisiCG} Bittner and Silva~\cite{Bittner14} and Ono and Ohno\cite{Ono16b}, but it can be applied to much larger systems. Thus, the model can provide a meaningful description of OPV operation at the mesoscale (10 nm or higher),\cite{Savoie14} which is crucial to account for the effect of blend morphology.
\textgx{Unlike Bittner and Silva,~\cite{Bittner14} the present version of our model does not account the coupled electron-nuclear dynamics, which are responsible for decoherence phenomena.
On the other hand,} several of the previous facilitators of CT dissociation may be readily introduced in a calculation, allowing a systematic and unbiased assessment of their relative importance.



\section{The model}

We model portions of a photoactive layer consisting of equal number of D and A sites. 
Overall there are $M=12\times 12\times 12=1728$ sites, arranged on a simple cubic lattice with a spacing of 1.0 nm. Figure S.1 in the Supporting Information shows the structure of our six model heterojunctions. The simplest one has a planar interface between the D and A sites, the others present some interpenetration between the phases to give a "comb" morphology (the systems have been named N$n$T$t$, according to the number $n$ of pillars and the thickness $t$ of the intermixing region).
In general, each site may represent a whole molecule, or a $\pi$-conjugated section of a long polymer chain.
There are two electrons and two orbitals per site, representing its highest occupied molecular orbital (HOMO) and lowest unoccupied molecular orbital (LUMO). 
This picture is similar to the one adopted in KMC simulations of OPV devices, but here the electronic states are derived from a proper quantum-mechanical description, without any assumption about their localization or delocalization.

The on-site parameters of our model Hamiltonian (the HOMO and LUMO energies, as well as the Coulomb and exchange interactions among the electrons) can be chosen to reproduce exactly the energies of the main electronic states of the individual sites/molecules: lowest singlet excitation energy ($SX$), lowest triplet excitation energy ($TX$), ionization energy to form a cation ($IE$) and electron affinity to form an anion ($EA$).
In our calculations, we employ a set of on-site parameters which correspond to C$_{70}$ for the acceptor 
and pentacene for the donor.\cite{refCGQC} 
Taking the gas-phase experimental data as a starting point, the effect of the surrounding dielectric on the ionized states has been obtained from the Born formula for the solvation free energy of an ion. All these energies have been collected in Table \ref{Tab1}.
Alternatively, this information about the single-molecule states could be obtained by conventional quantum chemical calculations, which can account for the surrounding dielectric by a polarizable continuum model\cite{PCM}.

The inter-site part of the Hamiltonian consists of a one-electron part, describing off-diagonal orbital couplings and the interaction with the positively charged cores of the other sites, and a two-electron part. The orbital couplings are assumed to be essentially random and decay exponentially with inter-site distance.
The inter-site electron-electron and electron-core interactions are approximated by the electrostatic interaction between two spherical Gaussian charge distributions, embedded in a dielectric medium with relative permittivity $\varepsilon_r=3.5$.
\textgx{As side note, we point out that situations with degenerate or near-degenerate HOMO and LUMO levels occur frequently in fullerene-based and other materials. These could be modelled within our coarse-grained model by connecting three or four sites into "super-molecules" with $D_{3h}$ or $T_{d}$ symmetry. Two sites could be considered to be connected when their coupling is roughly one order of magnitude larger than that between unconnected ones. Orbital degenerary might indeed have an effect on charge separation and transport,\cite{OD1,OD2} and we hope to study it in the future.}

\begin{table}
\begin{tabular}{|l|cc|cc|}
\hline
Energies & \multicolumn{2}{c|}{Vacuum} & \multicolumn{2}{c|}{Dielectric} \\
         &  A & D &  A & D \\
\hline
 $IE_r$ & 7.48 & 6.61 & 6.45 & 5.58 \\
 $EA_r$ & 2.68 & 1.35 & 3.71 & 2.38  \\
 $SX_r$ & 2.44 & 2.28 & 2.44 & 2.28 \\
 $TX_r$ & 1.56 & 1.76 & 1.56 & 1.76 \\
 \hline
\end{tabular}
\caption{Single-molecule energies (in eV) for the acceptor A (C$_{70}$) and the donor D (pentacene), in the gas phase or within a dielectric with $\epsilon_r=3.5$.
\label{Tab1}}
\end{table}

Both diagonal and off-diagonal disorder can be introduced in a controlled way, by admitting local deviations in the orbital energies and couplings. Our method allows independent, essentially unrestricted variations in their relative sizes.
Here we model them as random numbers, drawn from Gaussian distributions with standard deviations $\sigma_w=0.08$eV (for diagonal energetic disorder) and $\sigma_t=0.08$eV (for off-diagonal coupling disorder).
These values are comparable to those which arise in the calculation of charge mobilities in organic semiconductors.\cite{TroisiCT,ShuaiCT,Blumb}
For a given arrangement of D and A sites, different realizations of the disorder can be generated by simply re-assigning these energies and couplings, starting from a different random number seed. Typically, in order to extract systematic trends from our calculations, we consider one hundred independent realizations of the disorder.

The ground state energies, wavefunctions and charge distributions have been obtained by self-consistent-field, restricted Hartree-Fock (HF) calculations.\cite{mcweeny,schatz} The analogous excited state properties have been obtained by configuration interaction calculations including all the single excitations (CIS)\cite{cis} from the $M$ occupied to the $M$ virtual HF orbitals. CIS can be considered an excited-state extension of the HF method, as both of them neglect electron correlation effects. Note that the singly excited configuration do not contribute to the ground-state wavefunction, as a consequence of the variational nature of the HF solution (Brillouin's theorem).
\textgx{We have also evaluated the effect of non-dynamical electron correlation, by comparing the results of CIS and high-level excited-state calculations (equation-of-motion coupled-cluster singles plus doubles, or EOM-CCSD).\cite{refCC1,refCC2,refCC3} These results will be presented at the end of the following section.}
All calculations have been carried out with a modified version of GAMESS-US.\cite{refGMS} Further details are given in the Supporting Information.

We consider the lowest energy states obtained from the CIS calculations to be close theoretical relatives of the cold CT$_1$ states mentioned in the Introduction.~\cite{Salleo14,Neher14}
Our states are coherent or "pure", being described by a stationary electronic wavefunction instead of a density matrix.\cite{schatz}
The current version of the model does not deal with nuclear motions, and in particular with decoherence  and geometrical relaxation associated with charge transfer events.\cite{Scholes16,Koen}
\textgx{These phenomena have been modelled at the atomistic level by calculating explicitly the time-dependence of the nuclear coordinates and electronic wave function.\cite{OP14,Rocco15,Aki16} Clearly, this approach cannot be directly applied to the present model.
However, a time-dependent extension could be attempted by introducing a dependence of the orbital energies and couplings on some generalized intra- and intermolecular phonon coordinates.\cite{Tretiak,Bittner14}}
 
\section{Results and discussion}

As a preliminary step to the discussion of the electronic states at the heterojunctions, we provide some data on the pure materials. These have been obtained by HF and CIS calculations on blocks of $M$=1728 D-only or A-only sites. Considering for example the $i$-th realization of a donor block ($D_i$), the relevant energies are:
\begin{eqnarray}
SX(D_i) &=& E_{\mathrm{CIS}}(D_i) - E_{\mathrm{HF}}(D_i) , \nonumber \\
IE(D_i) &=& E_{\mathrm{HF}}(D_i^+) - E_{\mathrm{HF}}(D_i) , \label{Denergies} \\
EA(D_i) &=& E_{\mathrm{HF}}(D_i) - E_{\mathrm{HF}}(D_i^-) , \nonumber
\end{eqnarray}
where $E_{\mathrm{CIS}}$ is the energy of the first excited state of the neutral system, $E_{\mathrm{HF}}$ is the ground-state energy of the neutral or charged system.
Table \ref{Tab0} gives the averages and standard deviations of these energy differences, obtained by calculation on several independent realizations of the disorder.
The $SX$ energies of A and D compare favourably with the optical band gaps of solid C$_{70}$ (1.66 eV)\cite{C70gap} and pentacene (1.85 eV)\cite{PENTAgap}, respectively. At the same time, these energies are substantially lower than those of the single-molecule, on-site excitations (see again Table \ref{Tab1}). We take this as an indication that delocalization effects are significant and they are reasonably well described with our choice of the orbital coupling parameters.

\begin{table}
\begin{tabular}{|l|ccc|}
\hline
Material & $SX$ & $IE$ & $EA$ \\
\hline
 A & 1.62$\pm$0.17 &  5.99$\pm$0.21 &  4.17$\pm$0.15 \\
 D & 1.87$\pm$0.22 &  5.11$\pm$0.19 &  2.85$\pm$0.25 \\
\hline
\end{tabular}
\caption{Averages and standard deviations of the singlet excitation energies, ionization energies and electron affinities (in eV) of the pure materials.  \label{Tab0}}
\end{table}

The $IE$ and $EA$ data in Table \ref{Tab0} are also interesting, as they allow us to estimate the average energy of an electron-hole pair at infinite separation. Putting the hole on the donor and the electron on the acceptor, we find:
\begin{equation}
E_{eh}^\infty = IE(D)-EA(A) .
\end{equation}
The result is $E_{eh}^\infty=0.94\pm 0.24$ eV. This is roughly one half of the energy which could be estimated from the single-molecule data in Table \ref{Tab1}, assuming fully localized charges (1.87 eV). This confirms the importance of delocalization effects, despite of the sizeable amount disorder which has been included in the model.

\begin{figure}
\centering
\includegraphics[scale=0.35]{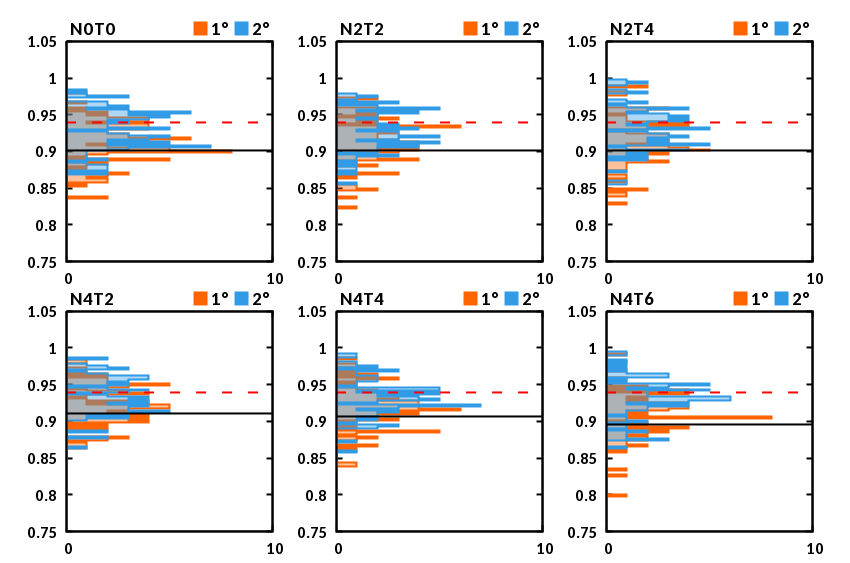} 
\caption{Distributions of the first and second excitation energies (in eV) for all the heterojunction morphologies. The horizontal axis is the number of counts within an energy bin. The horizontal black line marks the average first excitation energy. The dashed red line marks the average energy of an electron-hole pair at infinite separation ($E^\infty_{eh}$).
}
\label{Fig2}
\end{figure}

Figure \ref{Fig2} shows the distribution of the first and second excitation energies, calculated from one hundred independent realizations of each model heterojunction. Each of them might correspond to a slight rearrangement of the molecules, resulting from a combination of intra- and intermolecular vibrations. The lowest excitation energies are of the order of 0.90 eV. The average excitation energies (continuous black  lines in the Figure) do not seem to depend much on the interface morphology. These band gaps are substantially lower than those of the pure phases. At the same time, the CT$_1$ states are within only 0.04 eV from the average $E_{eh}^\infty$ (dashed red lines in the Figure). 
Note also that histograms of first and second excitations overlap strongly, each having a width of the same order of magnitude (comparable to $k_BT$). Thus, the second (or third) excitation energy of one system may be lower than the lowest excitation energy of another. In the following, we will concentrate on the discussion of the CT$_1$ states, but one should always keep in mind that there is actually a near-continuum of states above them.

We have just established that, using a resonable set of Hamiltonian parameters, the lowest interfacial excited states are within $k_BT$ from the infinitely separated charges, contrary to the "classical" expectations. 
\textgx{Before we look at this finding in greater depth, it is necessary to discuss its general validity. Clearly, by a proper choice of the Hamiltonian parameters, our calculations can be tuned to reproduce the vertical excitation and ionization energies of the don or and the acceptor.
Since the model does not account for geometrical relaxation following photoexcitation, our lowest excited states are "electronically cold" but "vibrationally hot", and we are certainly overestimating the transition energies that would be measured by fluorescence spectroscopy. On the other hand, the electron-hole pairs tend to be already well separated in these excited states (see below). It reasonable to assume that their vibrational relaxation energies are comparable to the sum of those of an electron within the A and a hole within the D. Thus, even though we are overestimating the energies of the relaxed interfacial excitons with respect to the ground state, we believe that the energies with respect to the fully separated charges should be roughly correct.}

\begin{figure}
\centering
\includegraphics[scale=0.35]{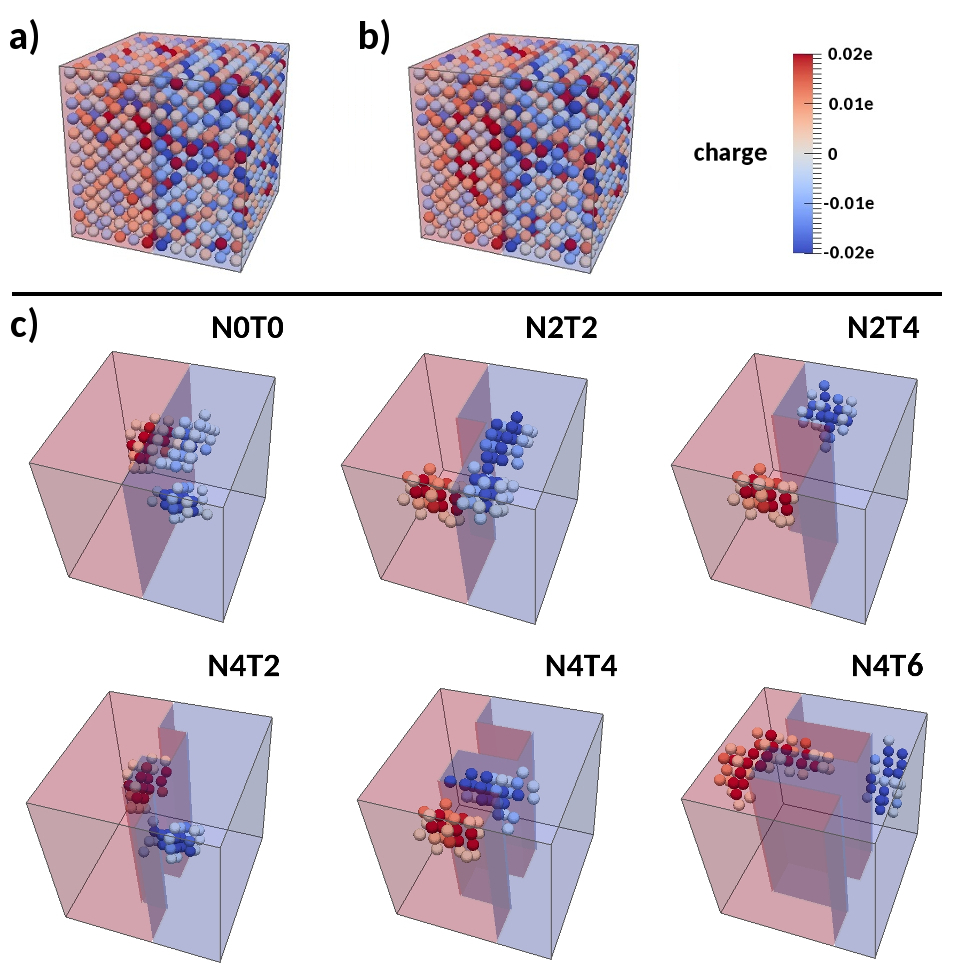} 
\caption{Upper panel: charge distributions in one realization of the bilayer, respectively for (a) ground state, (b) first excited singlet state. Lower panel: (c) charge differences between the excited and ground state, for one realization of each system. The red and blue shading of the background indicate the D and A sides of the heterojunctions, respectively, whereas the colouring of the sites indicates positive (red) or negative (blue) charges. For clarity, (c) shows only the sites with a charge difference greater than 0.002$e$ (in modulus).}
\label{Fig34}
\end{figure}

Figure \ref{Fig34} (upper panel) illustrates the charge distribution in the ground and the first excited states, for one realization of the flat bilayer. The sites are color-coded according to their net charges ($q_k^0$ for the ground state and $q_k^X$ for the excited state, $k$ being the site index). At first sight, the charges in the two states seem rather similar. In fact, they seem to be almost random, with several negative values on the D side and several positive ones on the A side of the interface. This randomness of the charge distribution reflects the randomness of the underlying Hamiltonian, of course.

A clear picture starts to emerge by plotting the charge difference between the states, $q_k^X-q_k^0$. Figure \ref{Fig34} (lower panel) shows several examples of the charge differences between the excited and the ground states, one for each instance of the model heterojunctions. The individual distributions are obviously rather different, but the overall picture is remarkably simple and consistent.
We see two distinct charge pockets, which are well separated and reside entirely on the expected phases (D for positive, A for negative). Summing these charge differences we always obtain:
\[     \sum_{k\in D}(q_k^X-q_k^0) = -\sum_{k\in A}(q_k^X-q_k^0) = 1.00e    \] 
so that photoexcitation produces a net transfer of one electron from the donor to the acceptor phase. These photogenerated charges appear to be delocalized over 10-20 sites, but this estimate can increase to 40-50 sites depending on the threshold adopted for their individual values.
Of course, this sizeable delocalization reflects the situation immediately following photoexcitation, before the nuclear motions set in to produce decoherence and further localization.


\begin{figure}
\centering
\includegraphics[scale=0.9]{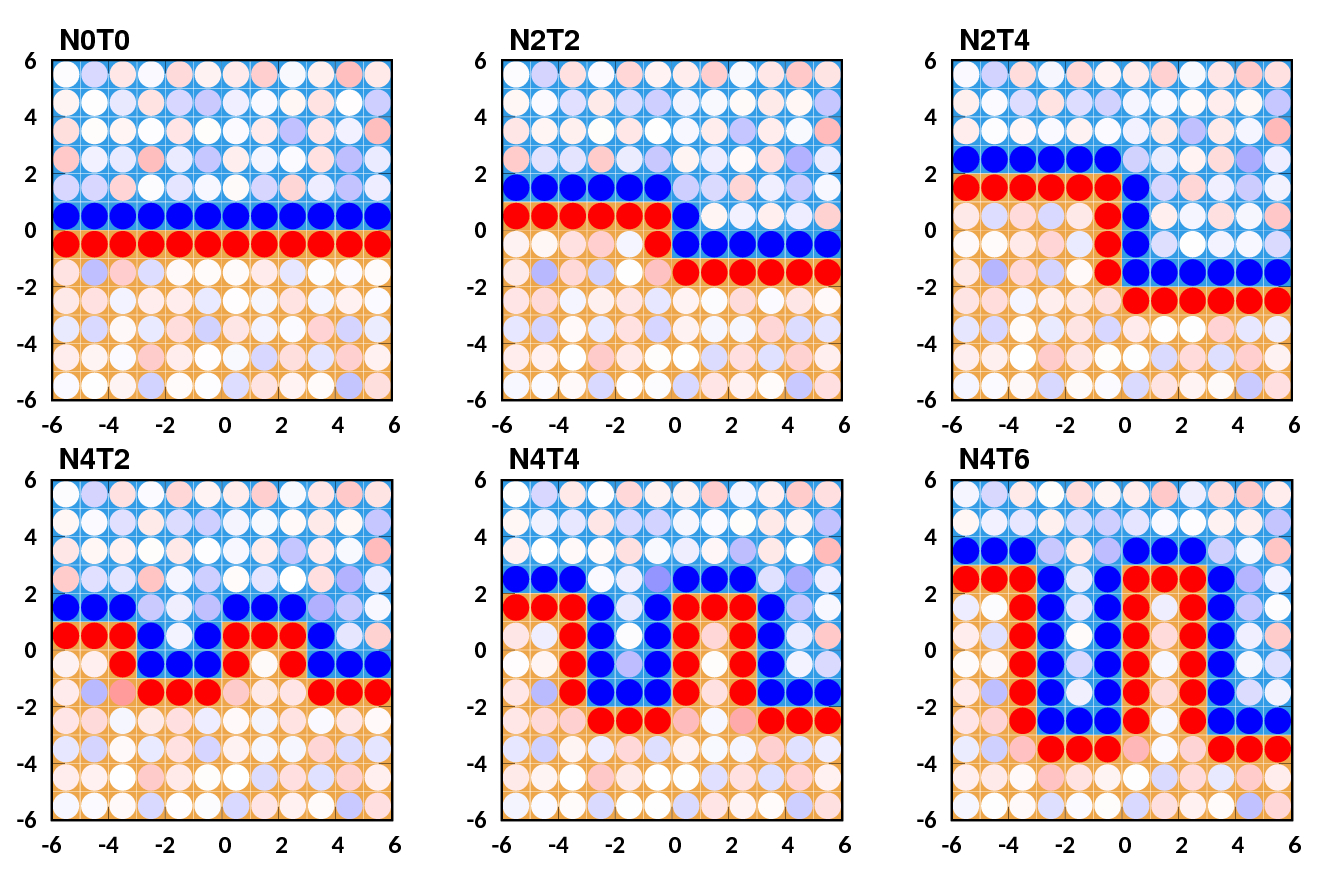} 
\caption{Two-dimensional plots of the average ground state charge distributions, $\langle q_k^0 \rangle$. The colour scale is such that saturation of the red and blue occurs when the charges exceed 0.0025$e$, in modulus.}
\label{Fig5}
\end{figure}

Because of the large variability in the charge distributions associated with different realizations of the Hamiltonian, it is necessary to average hundreds of them in order to extract further systematic trends. Figure \ref{Fig5} illustrates the results for the ground states. For clarity, we present them as simple two-dimensional maps. These have been extracted from the six layers at the center of the blocks, excluding those at the top and at the bottom (in the orientation of Fig. \ref{Fig34}) in order to minimize boundary effects. These charge density plots may also be taken to represent the result of an incoherent superposition of many quantum mechanical states, as all phase information is discarded when averaging their charge distributions. The ground state maps demonstrate the formation of sharp, well-defined interfacial dipoles. The average charge on a site sitting at the D/A interface is $0.007e$ (in modulus). Remembering that our sites are spaced by 1 nm, this is at least comparable with the interfacial charge densities of 0.02$e$/nm$^2$, as estimated by B\"{a}ssler and K\"{o}hler\cite{revBK} on the basis of the typical voltage drop at a heterojunction. The charges located on the sites away from the interface, although appreciable within a single realization of the disorder (see again Figure \ref{Fig34}), tend to cancel each other upon averaging.

\begin{figure}
\centering
\includegraphics[scale=0.9]{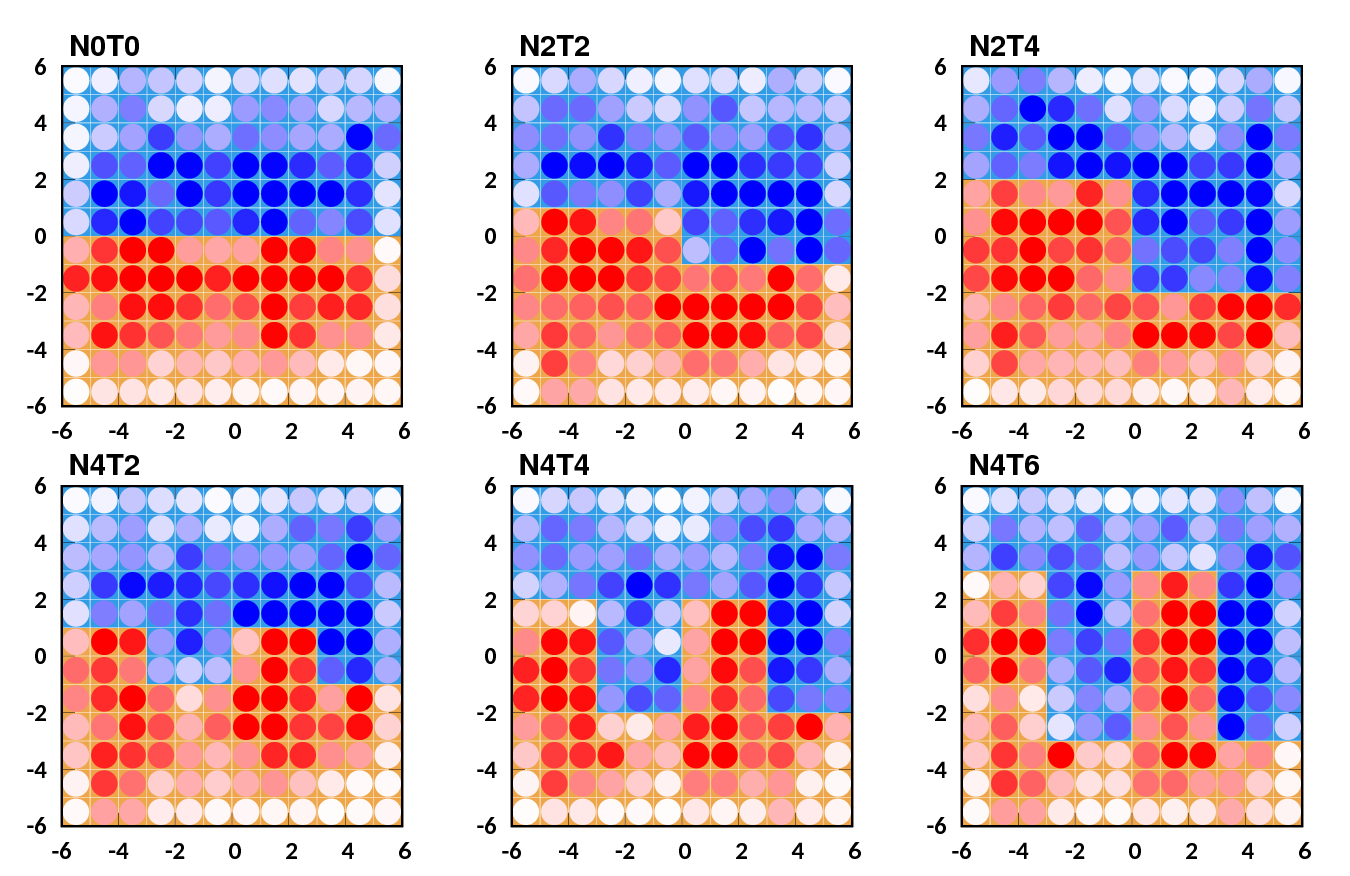} 
\caption{Two-dimensional plots of the average photogenerated excess charge distributions, $\langle q_k^X-q_k^0 \rangle$. The colour scale is identical to that of Figure \ref{Fig5}.}
\label{Fig6}
\end{figure}

Figure \ref{Fig6} contains the analogous two-dimensional maps of the photogenerated excess charge distributions. These plots tend to be noisier than those for the ground states, but even so it is possible to draw some interesting conclusions.
Naively, considering that we are looking at the lowest energy excitations, one would have expected the photogenerated charges to be "squeezed" at the interfaces by their mutual attraction. Instead, the average charge densities tend to be spread almost everywhere, except perhaps at the external boundaries of the blocks (good news, as this implies that finite-size effects are reasonably under control).
In particular, there is a significant photogenerated charge density within the first 2-3 layers from the interface.
\textgx{The subsequent time-dependent evolution will produce some localization of the photogenerated changes. Since we cannot simulate it with the current model, we can only speculate that their average distribution will still resemble the one in Figure \ref{Fig6}. In any case, the figure highlights that a truly unbiased model of charge photogeneration should include hundreds of donor and acceptor molecules, and this in currently incompatible with a detailed atomistic description of the coupled electron-nuclear dynamics.}

The electrostatic repulsion between the newly generated charges and the ground state interfacial dipoles---or, in other words, the built-in electric field---seems to be a key factor in enhancing this electron-hole separation. Exploratory calculations with other parameter sets show that an increased conjugation (introduced in the form of higher inter-obital couplings) and a lower disorder within the phases tend to produce an even greater electron-hole separation, up to a point where the photogenerated charge densities are almost zero on the D/A sites which are directly in contact. Thus, the calculations confirm the idea that both interfacial dipoles (a classical effect) and delocalization (a quantum-mechanical effect) can act as "facilitators" of charge dissociation.

\begin{figure}
\centering
\includegraphics[scale=1.0]{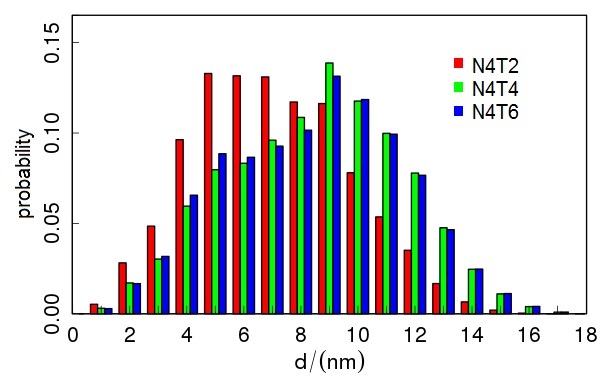}
\caption{Distribution of the electron-hole distances for selected interfacial morphologies.}
\label{Fig7}
\end{figure}

\textgx{Figure \ref{Fig7} provides further results about the effect of interfacial morphology on the distribution of electron-hole distances. Here the positions of the electrons and holes were identified by computing the center-of-charge of the excess photogenerated charges (negative and positive, respectively). The plots show that increasing the interfacial thickness from two (N4T2) to four or six nm's (N4T4 and N4T6) should have a beneficial effect on charge separation, by shifting the distribution of electron-hole distances toward larges values. If confirmed, this observation could turn into a useful design rule for the choice of photovoltaic materials and their assembly within the active layer. In the future, we hope to corroborate it by further calculations based on more realistic off-lattice models with "arctan-like" concentrations of the donor and the acceptor.}

\begin{figure}
\centering
\includegraphics[scale=0.2]{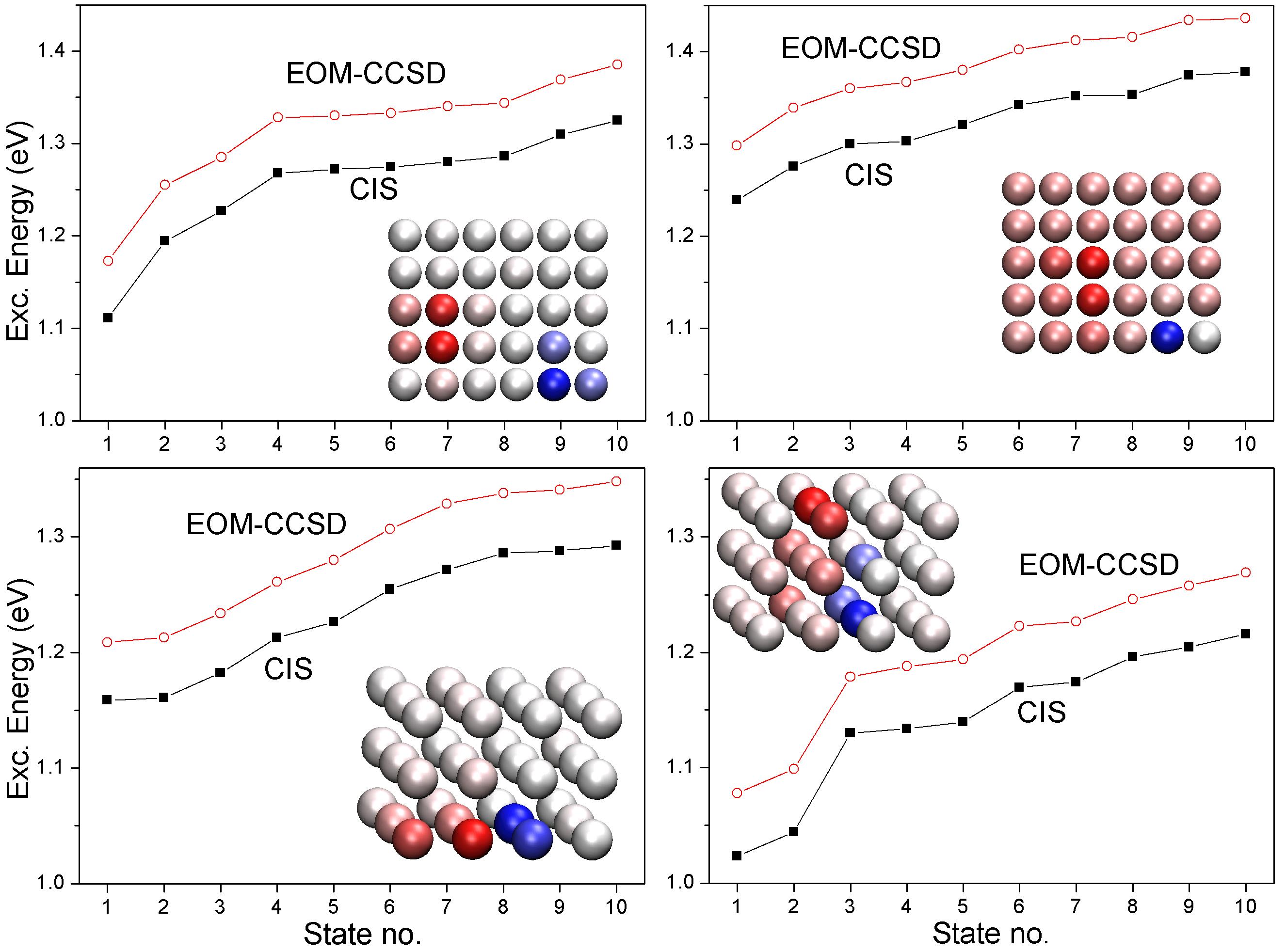} 
\caption{Comparison of CIS and EOM-CCSD calculations of the first ten excited states, in two-dimensional (above) and three-dimensional (below) model heterojunctions. The insets show the EOM-CCSD charge distributions in the first excited states.}
\label{Fig8}
\end{figure}

\textgx{Finally, we validate the use of the CIS method by comparing it with EOM-CCSD excited state calculations. We have carried out benchmark calculations on two systems: A two-dimensional heterojunction consisting of 15 D and 15 A sites, and a three-dimensional heterojunction consisting of 18 D and 18 A sites. Figure \ref{Fig8} shows the results for two independent realizations of each system.
In comparison with EOM-CCSD, CIS systematically underestimates the excitation energies by about 0.05 eV. This applies to all the lowest excited states. As a results, one excitation spectrum is simply a shifted version of the other. Note also the significant changes in the excitation spectra, produced by simply changing the random seed for the assignment of the orbital couplings. Once more, this confirms the importance of performing extensive statistics before drawing any conclusions. The insets in Figure \ref{Fig8} show also the EOM-CCSD charge distribution in the lowest excited state of each system. In all cases, the CIS charge distributions (not shown) are virtually indistinguishable from them. Thus, the neglect of electron correlation in the description of the excited states of these systems appears to have solid foundations. This is certainly good news also for those performing atomistic (ab initio or semiempirical) excited-state calculations.}

\section{Conclusions}

In conclusion, we have presented the results of mesoscale quantum mechanical calculations of the lowest excited states at organic photovoltaic interfaces. Despite of its relative simplicity---or perhaps thanks to it---the model allows an unbiased exploration of the effect of several possible facilitators of charge photogeneration, such as built-in or external electric fields, charge delocalization, disorder, nanostructuring of the donor-acceptor blends. The nature of the problem requires large numbers of calculations on large systems, as it is necessary to avoid assumptions about the localization of the photogenerated charges, minimize finite-size effects and perform adequate statistical averaging of the results. We have explored only a small fraction of the parameter space, for example using C$_{70}$ as the only acceptor and pentacene as the only donor. Nonetheless, the results are already very encouraging, as they show that fully separated charges (electron and hole) are easily within 1-2 $k_BT$ from the lowest excited states, in agreement with much experimental evidence but contrary to simple "classical" arguments. \textgx{Delocalization emerges from our calculations as one of the key facilitators of charge dissociation. We find that a significant degree of delocalization of the excited charge transfer states (tens of molecules/sites) may occur also in presence of some energetic and coupling disorder. On the other hand, nuclear relaxation phenomena are not included in our treatment, and as a result the real situation might be somewhere in between our description and the fully localized one which is assumed in classical KMC simulations of OPV devices.}

Our model is somewhat generic, as it retains a minimum of physically important features and neglects most molecular details. In this sense, it is analogous to many classical coarse-grained models of soft materials (polymers, liquid crystals, colloids, etc.), in which atomic level details are sacrificed in favour of generality, lower computational cost and greater interpretability of the results. Considering the relative ease of achieving $>90\%$ IQE's (relative to the difficulty of achieving high charge mobilities in organic semiconductors, for example), it seems that the photogeneration of charges is precisely a situation where a somewhat generic explanation is required. 
\textgx{This contrasts with other aspects of the behaviour of organic photovoltaic materials. Compare it, for example, with the difficulty of achieving high charge mobilities in molecular or polymeric semiconductors. This has eventually been possible thanks to a lengthy and painstaking selection of molecules and processing conditions, providing the optimal combinations of molecular structure and supramolecular organization. Clearly, the latter is a situation where detailed molecular-level modelling is necessary in order to support and guide the experimental efforts.
In any case, our calculations provide guidelines also for those performing more conventional, atomistic excited state calculation at D/A heterojunctions. There is a clear need to move towards models incorporating hundreds of molecules in order to remove any bias about the degree of localization. This is extremely challenging, considering also the need to average over many disordered configurations, but at least our coupled-cluster calculations show that these calculations should not require the inclusion of dynamical electron correlation.}

\textgx{We conclude with a brief perspective on future developments.} Although here we have considered idealized on-lattice models of the blends interfaces, more realistic off-lattice models are fully within our reach. It should also be interesting to examine the consequences of a polymer-like connectivity of donor or acceptor sites, to form long conjugated chains. 
\textgx{Bittner and Silva\cite{Bittner14} have already considered a two-dimensional system in which the both the donor and acceptor are polymeric. In their lattice model, all the chains are parallel to the interface and this produces low-energy charge transfer states in which the electron and hole are delocalized but "pinned" to the interface. However, the situation in a real system will be  more complicated, and even a partial orientation or $\pi$-stacking of the chains in the orthogonal direction might have a strong beneficial effect on charge separation.
In the longer term, the present model should be extended by incorporating electron-phonon coupling and nuclear relaxation effects, within an explicitly time-dependent picture of system. One day, with further developments along the lines of the present study, it might even become possible to simulate the operation of whole OPV devices by adding a quantum mechanical description of exciton diffusion, charge transport and charge extraction/injection at the electrodes.}

\textbf{Acknowledgements.}
The research of GR and MC is supported by the PRIN project "Molecular organization in organic thin films via computer simulation of their fabrication processes" (2015XJA9NT\_003).
The research of RaP is supported by the SPIN SEED 2014 project "Charge separation and
charge transport in hybrid solar cells" and the CNR--NTU joint laboratory "Amorphous materials for energy harvesting applications".

\textbf{Supporting Information} 
Detailed description of the theory and computational methods. References for the experimental gas-phase energies of pentacene and C$_{70}$. Illustration specifying the structure of the model heterojunctions. \\
This material is available free of charge at \texttt{http://pubs.acs.org/doi/abs/10.1021/acs.jpcc.7b03640}.

\end{document}


\maketitle

\section*{Theoretical details}

Our effective two-orbital quantum chemical model\cite{refCGQC} is based on the second-quantized Hamiltonian:
\begin{equation}
\hat{H} = \sum_{i,j=1}^{2M}\sum_{\sigma=\alpha}^{\beta} h_{ij} a_{i\sigma}^\dagger a_{j\sigma} +
          \frac{1}{2} \sum_{i,j,k,l=1}^{2M} \sum_{\sigma,\tau=\alpha}^{\beta}c_{ikjl} a_{i\sigma}^\dagger a_{j\tau}^\dagger a_{l\tau}a_{k\sigma} \label{hamiltonian}
\end{equation}
where $M$ is the total number of sites and $a_{i\sigma}^\dagger$ and $a_{j\tau}$ are electron creation and annihilation operators satisfying the usual Fermion anticommutation rules.\cite{mcweeny,schatz} 
Indices $i,j,k,l$ run over the orbitals (the HOMO  and LUMO on site $r$ are $2r\!-\!1$ and $2r$, respectively), $\sigma$ and $\tau$ over the possible spin states.
The orbitals $\phi_k$, which provide the basis for the second-quantized Hamiltonian, are assumed to be orthonormal:
\begin{equation}
\int \! \phi_{i}(\mathbf{r}) \phi_{j}(\mathbf{r}) d\mathbf{r} = \delta_{ij} .
 \label{Sij}
\end{equation}
The one-electron ($h_{ij}$) and two-electron integrals ($c_{ikjl}$) which form the Hamiltonian are formally defined as:
\begin{eqnarray}
 h_{ij} \!\!&=& \!\! \int \! \phi_{i}(\mathbf{r}) \left[ -\frac{1}{2}\nabla^2 - \sum_{p=1}^M \frac{Z_p^{\mathrm{eff}}}{|\mathbf{r}-\mathbf{R}_p|} \right] \phi_{j}(\mathbf{r}) d\mathbf{r} \label{1ei} \\
 c_{ikjl} \!\!&=& \!\!  \int\!\int\!
 \frac{\phi_{i}(\mathbf{r}_1) \phi_{k}(\mathbf{r}_1) \phi_{j}(\mathbf{r}_2) \phi_{l}(\mathbf{r}_2) } {|\mathbf{r}_1-\mathbf{r}_2|} 
                d\mathbf{r}_1 d\mathbf{r}_2 \, , \label{2ei}
\end{eqnarray}
where $\mathbf{R}_p$ is the location of site $p$.

We compute the electronic energies and wavefunctions of the model systems with a modified version of the GAMESS-US code.\cite{refGMS}  Within the program, each site is treated as a helium atom with a double-$\zeta$ basis. This ensures that that all the arrays are dimensioned correctly. Our version of the code bypasses the usual \textit{ab initio\/} evaluation of the one- and two-electron integrals from the orbitals, replacing them by semiempirical values chosen to account for the essential physics of a system. This approach is closely analogous in spirit to the early semiempirical theories of molecular electronic structure and to the Hubbard model of solid-state physics. The remainder of our code is essentially identical to the standard version of GAMESS-US. One ground-state (HF) plus excited-state (CIS) calculation on one of our systems takes about 15 minutes, on a single-processor workstation. The use of GAMESS-US also allows us to leverage on the full range of quantum chemical methods implemented in it. For example, in addition to the HF and CIS calculations\cite{cis} presented within the paper, it is possible to perform ground- and excited-state calculations based on the coupled-cluster \textit{ansatz}.\cite{CC} 

The on-site parameters of the Hamiltonian for a molecule $r$ are related to its ionization energy ($IE_r$), electron affinity ($EA_r$), singlet excitation ($SX_r$) and triplet excitation ($RX_r$):
\begin{equation}
\left[ \begin{array}{c}
IE_r \\ EA_r \\ SX_r \\ TX_r
\end{array} \right] = 
\left[ \begin{array}{rrrr}
 -1 &  0 & -1 &  0 \\
  0 & -1 & -2 &  1 \\
 -1 &  1 &  0 &  1 \\
 -1 &  1 &  0 & -1 
\end{array} \right]
\left[ \begin{array}{c}
 \epsilon^H_{r} \\ \epsilon^L_{r} \\ c^C_{r} \\ c^X_{r}
\end{array} \right]                                     \label{params1}
\end{equation}
where $\epsilon^H_{r}$ and $\epsilon^L_{r}$ are the HOMO and LUMO energies of site $r$, while $c^C_{r}$ and $c^X_{r}$ are short-hand notations for the two-electron integrals describing on-site Coulomb and exchange interactions ($c^C_{r} = c_{2r-1,2r-1,2r-1,2r-1}\simeq c_{2r-1,2r-1,2r,2r}\simeq c_{2r,2r,2r,2r}$ and $c^X_{r}=c_{2r-1,2r,2r-1,2r}$).
Equating the three on-site Coulomb integrals is a resonable and convenient approximation, which could be easily avoided if it were necessary to reproduce some additional single-site energies (e.g., the singly excited states of the cation and anion, or the doubly excited state of the neutral molecule).
Inverting Eq. (\ref{params1}) we obtain the Hamiltonian parameters as a function of the energies:
\begin{equation}
\left[ \begin{array}{c}
 \epsilon^H_{r} \\ \epsilon^L_{r} \\ c^C_{r} \\ c^X_{r}
\end{array} \right] = 
\left[ \begin{array}{rrrr}
 -2 &  1 &  0 &  1 \\
 -2 &  1 & 1/2 & 3/2 \\
  1 & -1 &  0 & -1 \\
  0 &  0 & 1/2 & -1/2
\end{array} \right]
\left[ \begin{array}{c}
IE_r \\ EA_r \\ SX_r \\ TX_r
\end{array} \right]  .                                   \label{params2}
\end{equation}

The Coulomb integrals $c^C_{r}$ provide a rough estimate of the spatial extent of the  orbitals associated with a site. Let us assume that one electron within an orbital produces a Gaussian charge distribution:
\begin{equation}
\rho_r(\mathbf{r}) = - \phi_r^2(\mathbf{r}) = - \left(\frac{2\alpha_r}{\pi}\right)^{3/2} e^{-2\alpha_r|\mathbf{r}-\mathbf{R}_r|^2}    \label{gaussian}
\end{equation}
Its self-repulsion integral is:
\begin{equation}
c^C_{r} = \int\int \frac{\rho_r(\mathbf{r}_1)\rho_r(\mathbf{r}_2)}{|\mathbf{r}_1-\mathbf{r}_2|}d\mathbf{r}_1d\mathbf{r}_2 
         = \sqrt{\frac{4\alpha_r}{\pi}} \, .    \label{width}
\end{equation}
Reversing this equation, we obtain the standard deviation $\sigma_r$ from the integral:
\begin{equation}
\sigma_r = \sqrt{\frac{3}{4\alpha_r}} = \sqrt{\frac{3}{\pi}}\frac{1}{c^C_{r}} \, .  \label{sigma}
\end{equation}
Using the Coulomb integrals resulting from Eq. (\ref{params2}) and the gas-phase data of C$_{70}$\cite{C70refa,C70refb,C70refc} and pentacene\cite{PENTArefa,PENTArefb,PENTArefc,PENTArefd} (see also Table 1 within the main manuscript), we obtain $\sigma_r=0.434$ nm for the former and $\sigma_r=0.402$ nm for the latter. These values correspond to roughly one half of the sites' diameters.

The ionization energies and electron affinities which we use in our calculations are derived from the experimental gas phase ones, after adjusting for polarization effects. Our estimate of this correction is based on the Born formula for the solvation free energy of a spherical charge $\pm e$ of diameter $D_0=1.0$ nm (i.e., the nearest-neighbour distance) inside a dielectric with relative permittivity $\varepsilon_r=3.5$:
\begin{equation}
\Delta G_{\mathrm{Born}} = - \frac{e^2}{D_0} \left( 1-\frac{1}{\varepsilon_r}\right) \, .  \label{Born}
\end{equation}
This lowers the energies of both the anion and cation states, decreasing the $IE$ and increasing the $AE$ by about 1 eV (see again Table 1 within the main manucript).


We now consider the Hamiltonian for many interacting sites.
First of all, the orbital energies on one site are shifted by the interaction with the cores of the other sites. We assume that the charge distribution of one core is the positive, doubly charged version of its HOMO density (see Eq.(\ref{gaussian})):
\begin{equation}
\rho^{\mathrm{core}}_r(\mathbf{r}) = +2\phi_r^2(\mathbf{r}) = +2 \left(\frac{2\alpha_r}{\pi}\right)^{3/2} e^{-2\alpha_r|\mathbf{r}-\mathbf{R}_r|^2}    \label{gaussian2}
\end{equation}
The on-site elements of the one-electron Hamiltonian become:
\begin{eqnarray}
h_{2r-1,2r-1} &=& \epsilon_r^H + w_{2r-1} - \sum_{p\neq r}^M  \frac{ 2 \mathrm{erf}\!\left( \mu_{rp}R_{rp} \right)}{\varepsilon_r R_{rp} } \label{h11} \\
h_{2r,2r} &=& \epsilon_r^L + w_{2r} - \sum_{p\neq r}^M \frac{ 2 \mathrm{erf}\!\left( \mu_{rp}R_{rp} \right)} {\varepsilon_r R_{rp} } \label{h22} \\
h_{2r-1,2r} &=& 0
\end{eqnarray}
where erf(x) is the error function and $\mu_{rp}=\sqrt{2\alpha_r\alpha_p/(\alpha_r+\alpha_p)}$.
The $w$'s in Eqs. (\ref{h11}) and (\ref{h22}) are perturbations to the site energies, which model the effect of "diagonal" or "energetic" disorder.  We draw these numbers from Gaussian distributions of width $\sigma_w=0.08$ eV, assumed for simplicity to be identical for both D or A materials.

Next, we assume that the inter-site elements of the one-electron Hamiltonian decay exponentially with the distance $R_{ij}$:
\begin{equation}
h_{ij}=t_{ij} e^{-(R_{ij}-D_0)/\Delta} .   \label{hij}
\end{equation}
Here $t_{ij}$ represents the coupling between two orbitals at the nearest-neighbour distance $D_0$, and $\Delta$ determines the decay of the couplings with increasing separation. A controlled degree of "off-diagonal" disorder can be introduced by assuming that the $t_{ij}$'s are drawn from suitable distributions. We assume these to be Gaussians. In principle, both their averages and their widths may depend on the materials and on the orbital types. In the absence of further information, coming for example from detailed atomistic models of the individual materials and their interfaces, we assume that all the couplings are symmetrically distributed around a zero average with a standard deviation $\sigma_t$=0.08 eV (identical to $\sigma_w$). Furthermore, we take $\Delta=0.35$ nm.

The inter-site electron-electron repulsions are represented by two-electron integrals. All three- and four-center integrals are neglected, and we retain only the two-center integrals which represent the classical repulsion between two Gaussian charge clouds:
\begin{equation}
c_{ikjl} = \delta_{ik} \delta_{jl} \frac{ \mathrm{erf}\!\left( \mu_{ij}R_{ij} \right) }{\varepsilon_r R_{ij} } . \label{2eints}
\end{equation}
Thanks to this zero-overlap approximation, the number of two-electron integrals to be processed in a calculation is substantially smaller than in an \textit{ab initio} calculation with a comparable basis set.

The dipole integrals, which are necessary to compute the dipole moments and to study the effect of an external electric field, are approximated in a way consistent with the zero-overlap approximation:
\begin{equation}
 \mathbf{\mu}_{ij} = \int \! \phi_{i}(\mathbf{r}) \mathbf{r} \phi_{j}(\mathbf{r}) d\mathbf{r}
                   \approx \delta_{ij} \mathbf{R}_i
 \label{muij}
\end{equation}
where $\mathbf{R}_i$ is the location of orbital $i$. Note that the dipole integral between the HOMO and LUMO on the same site is also zero, in this approximation.

The site charges are obtained by conventional Mulliken or L\"{o}wdin populational analyses, the two being equivalent in the case of orthogonal orbitals.\cite{mcweeny} With the dipole integrals of Eq.(\ref{muij}), the quantum mechanical dipole moments, calculated as expectation values of the ground or excited state wavefunctions, coincide with the "classical" ones for a set of point charges $q_k^S$ at the sites with coordinates $\mathbf{R}_k$ (the $S$ superscript identifies a state):
\begin{equation}
 \mathbf{\mu}^S = \sum_{k=1}^M q_k^S \mathbf{R}_k . \label{muS}
\end{equation}

\begin{figure}
\centering
\includegraphics[scale=0.30]{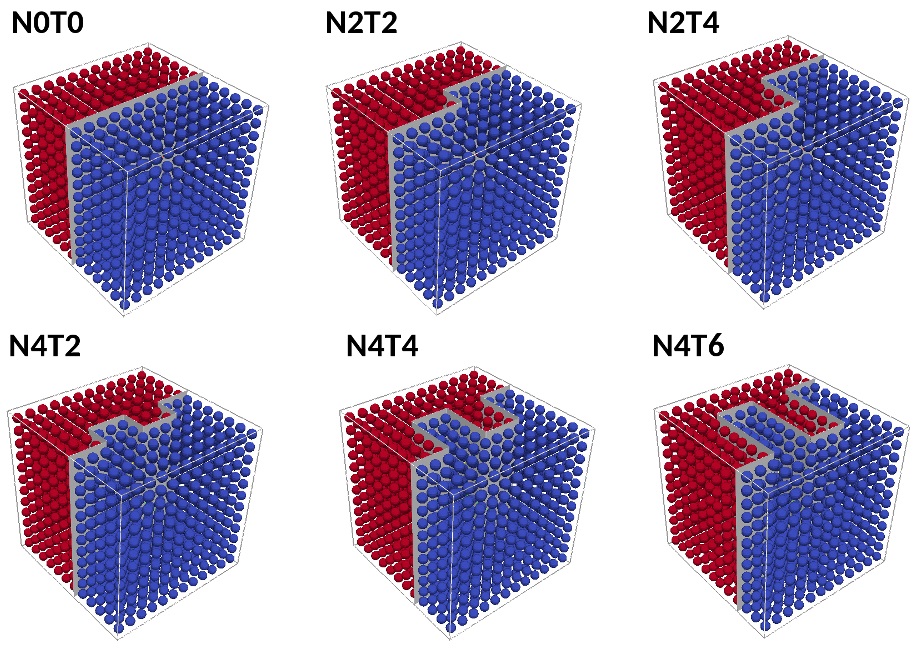} 
\caption{Illustration and denomination of the three-dimensional model heterojunctions. Donor sites are depicted in red, acceptor sites in blue.}
\end{figure}

\section*{Interface models}

Figure S.1 shows the structure of all the three-dimensional model heterojunctions. The simplest one (N0T0) is a bilayer with a planar interface between the D and A sites, orthogonal to the $z$ axis. We have also studied the effect of variations on this basic system, introducing some interpenetration between the two phases in the form of a "comb" morphology. The systems have been named according to the thickness of the mixing region (T, in number of D:A layers) and to the number of pillars (N).